\documentclass[12pt,preprint]{aastex}
\usepackage{natbib}
\usepackage{lscape}
\shorttitle{First VLF detection of magnetar short bursts}
\shortauthors{Tanaka et al.}

\begin{document}


\title{First VLF detection of short repeated bursts from magnetar SGR J1550$-$5418}


\author{Y. T. Tanaka\altaffilmark{1}, Jean-Pierre Raulin\altaffilmark{2}, Fernando C. P. Bertoni\altaffilmark{2},
P. R. Fagundes\altaffilmark{3}, J. Chau\altaffilmark{4}, N. J. Schuch\altaffilmark{5},
M. Hayakawa\altaffilmark{6}, Y. Hobara\altaffilmark{6},
T. Terasawa\altaffilmark{7}, and T. Takahashi\altaffilmark{1,8}}
\email{tanaka@astro.isas.jaxa.jp}

\altaffiltext{1}{Institute of Space and Astronautical Science, Japan Aerospace Exploration Agency, Japan.}
\altaffiltext{2}{CRAAM, Presbyterian Mackenzie University, S\~{a}o Paulo, Brazil.}
\altaffiltext{3}{University of Vale do Paraiba, S\~{a}o Jos\'{e} dos Campos, Brazil.}
\altaffiltext{4}{Radio Observatorio de Jicamarca, Instituto Geof\'{i}sico del Per\'{u}, Lima, Per\'{u}.}
\altaffiltext{5}{INPE's Southern Regional Space Research Center, Santa Maria, Brazil.}
\altaffiltext{6}{University of Electro-Communications, Japan.}
\altaffiltext{7}{Institute for Cosmic Ray Research, University of Tokyo, Japan. }
\altaffiltext{8}{University of Tokyo, Japan.}

\begin{abstract}
We report on the first detection of ionospheric disturbances caused by {\it short repeated} gamma-ray bursts from the magnetar
SGR J1550$-$5418. Very low frequency (VLF) radio wave data obtained in South America clearly show
sudden amplitude and phase changes at the corresponding times of eight SGR bursts.
Maximum amplitude and phase changes of the VLF signals appear to be correlated with the gamma-ray fluence. On the other hand,
VLF recovery timescales do not show any significant correlation with the fluence, possibly suggesting that the bursts' spectra are not 
similar to each other.
In summary, the Earth's ionosphere can be used as a very large gamma-ray detector and
the VLF observations provide us with a new method to monitor high energy astrophysical phenomena without
interruption such as Earth Occultation. 

\end{abstract}





\section{Introduction}
Very low frequency (VLF; 3$-$30 kHz) radio waves are reflected at the Earth's lower ionosphere and ground, and
propagate within the Earth-ionosphere waveguide \cite[e.g.,][]{Wait1962}.
Since amplitude and phase of VLF radio waves are sensitive to the condition of the
lower ionosphere, they have been utilized to investigate the physics of the lower ionosphere.
Energetic electrons can precipitate into the ionosphere due to wave-particle interaction in the magnetosphere and
cause VLF signal amplitude and phase perturbations \cite[e.g.,][]{Kikuchi1983}.
Soft X-rays from solar flares are also another source of ionospheric disturbances, which are detected using VLF signals
\cite[e.g.,][]{Todoroki2007, Raulin2010}.

Besides these solar-terrestrial events, the lower ionosphere is also affected by high energy photons
(X-rays and gamma-rays) from extra-solar sources. An ionospheric disturbance caused by a cosmic gamma-ray burst
was first reported by \citet{Fishman1988}. It suggested that gamma-rays deposit their energies
in the lower ionosphere, ionize abnormally the neutral atmosphere there, and modify the electron density height profile.
In addition, it is known that giant flares from Soft Gamma-ray Repeaters (SGRs, also called magnetars) 
significantly affect the lower ionosphere \citep{Inan1999, Inan2007, Tanaka2008}.

Magnetars emit a lot of short-duration gamma-ray flares repeatedly during active phases \cite[e.g.,][]{Woods2006}.
Typical duration and flux of the short repeated bursts are 0.1$-$1 s and 10$^{-6}$ erg s$^{-1}$ cm$^{-2}$, respectively.
Furthermore, magnetars rarely emit exceptionally bright gamma-ray flares (giant flares).
So far, only three giant flares were recorded. The first one was detected in 1979, from the source SGR 0526$-$66 \citep{Mazets79}.
The second and third ones were emitted by SGR 1900+14 and SGR 1806$-$20, and they were observed by satellites in 1998 \citep{Hurley1999, Mazets1999, Tanaka2007ApJ} 
and 2004 \citep{Terasawa2005, Hurley2005, Palmer2005, Frederiks2007}, respectively.
Since the fluences were much larger than those of GOES X-class solar flares by a few orders of magnitude, ionospheric disturbances
caused by these giant flares were detected as sudden and large amplitude changes of VLF radio waves \citep{Inan1999, Inan2007, Tanaka2008}.

On the other hand, VLF amplitude and phase changes caused by {\it short repeated} gamma-ray bursts from a magnetar have not been detected so far
because of the lack of high sensitivity of VLF observing systems. 
On 22 January 2009, one of the known magnetars SGR J1550$-$5418 emitted a lot of short-duration gamma-ray bursts repeatedly \citep{Mereghetti2009}. 
In this paper, we report on the first VLF detection of short repeated gamma-ray flares from this object.
VLF data were provided by the South America VLF Network (SAVNET) tracking system \citep{Raulin2009}. In \S2, we describe details of the SAVNET observations.
Comparison of VLF amplitude and phase changes with gamma-ray fluences measured by {\it INTEGRAL} satellite are presented and discussed in \S3 and \S4.
We summarize this paper in \S5.

\section{Observations}
The VLF data shown in this paper were obtained by SAVNET, which was recently installed
in Brazil, Peru, and Argentina (see \citet{Raulin2009} for the details of the SAVNET instrumental facility).
Figure 1 shows locations of a relevant observing station ATI (Atibaia, S\~{a}o Paulo, Brazil) as well as five VLF transmitters (NPM, NLK, NDK, NAU, and NAA),
and VLF signals from them have been continuously recorded with the time resolution of 1 sec. 
The propagation path from NPM (21.4 kHz) to ATI is also drawn in Figure 1.
Shaded hemisphere in Figure 1 exhibits the night-side part of the Earth at 6:48 UT, when the most intense gamma-ray flare
occurred. The point on the Earth directly beneath the flare (subflare point) was located at
54.3$^{\circ}$S, 14.0$^{\circ}$E, and its position is shown using a cross.
The part of the Earth illuminated by gamma-rays at 6:48 UT is illustrated by dashed area.

Figure 2 shows NPM-ATI amplitude and phase data recorded from 4:00 UT to 10:00 UT on 22 January, 2009.
In Figure 3 we also display an extended view of NPM-ATI data together with the gamma-ray light curve\footnote{
http://www.isdc.unige.ch/integral/ibas/cgi-bin/ibas\_acs\_web.cgi}
observed by {\it INTEGRAL} satellite around 6:48 UT \citep{Mereghetti2009}.
Due to the high sensitivity of the SAVNET facility, we can clearly see rapid amplitude and phase changes
at the corresponding times of the short repeated bursts from SGR J1550$-$5418.
Therefore, we can robustly claim that the rapid changes were caused by the short gamma-ray bursts from the magnetar. 
We listed in Table 1 the properties of the SGR short bursts detected by the NPM-ATI VLF propagation path. 
Although these magnetar bursts were detected at other SAVNET receiving stations, like PAL (Palmas, TO, Brazil), 
SMS (S\~{a}o Martinho da Serra, RS, Brazil), PIU (Piura, Peru) and EACF (Antarctica), in this short paper
we concentrate on the records from ATI receiving station. A detailed comparison of VLF phases and amplitudes observed by the other receivers
is out of the scope of this paper, and will be reported in a subsequent forthcoming article.

\section{Analysis}
To investigate the influence of gamma-ray irradiation in the lower ionosphere, 
we need to characterize the VLF amplitude and phase data.
To do this, we estimated the maximum amplitude ($\Delta A$) and phase variations ($\Delta \phi$) as well as the recovery timescales 
of the amplitude as follows.
First, we extracted the data before and after each burst, and chose a proper functional form to represent the baseline level.
Most of the baseline levels can be well fitted by first-order polynomial functions.
When the baseline levels showed curvature, we used second-order polynomial functions to represent them.
After subtracting the trends from the amplitude and phase data, we obtained $\Delta A$ and $\Delta \phi$, which
are tabulated in Table 1. To estimate the typical errors of $\Delta A$ and $\Delta \phi$,
we made histograms of the residuals, which were distributed around 0 with a Gaussian-like form.
Therefore, we fitted the histogram using a Gaussian, and took the variance as a typical error.

To quantify the recovery timescales of the VLF amplitude data, 
we have used the function
\begin{equation}
f(t)=\left( {\rm Baseline} \ {\rm level} \right) - \\
\frac{F_0}{\left\{ \exp((t_0 - t)/t_{\rm fall}) + \exp((t - t_0)/t_{\rm rcv}) \right\}},
\end{equation}
where $F_0$ is a typical amplitude decrease, $t_{\rm fall}$ is a falling time, and $t_{\rm rcv}$ is a
recovery timescale. We fitted the data using this function, and determined $t_{\rm rcv}$ for each burst (see Table 1).

\section{Discussion}
\subsection{Amplitude and phase changes}
In Figures 4 (a) and 4 (b), we plot $\Delta A$ and $\Delta \phi$
against the gamma-ray fluence from 25 keV to 2 MeV for each burst (see also Table 1).
Although there are not many data points, possible correlations between $\Delta A$ and $\Delta \phi$, and gamma-ray fluences are seen.
We note that similar correlations were also reported in the case of X-ray solar flares \citep{Mcrae2004,Pacini2006}

We can understand these evidences of correlation in terms of the lowering of the reflection height due to gamma-ray ionization.
Under a typical undisturbed nighttime condition, VLF waves are thought to be reflected at $\sim$85 km \citep[e.g.,][]{Carpenter1997}.
When gamma-rays are directed onto the Earth, they deposit most of their energy in the lower ionosphere, ionize the neutral atmosphere there, and
produce free electrons. The typical altitude where these free electrons are produced depends on the photon energy.
For example, by using Monte Carlo simulation \citet{Inan1999} reported that 3 keV and 10 keV photons mainly ionize the atmosphere
at $\sim$82 km and 60 km, respectively. Similar calculations have shown that 
gamma-ray illumination increases electron number density below $\sim$85 km, depending on the photon spectrum \citep[e.g.,][]{Brown1973, Baird1974, Tanaka2008}.
Consequently, the VLF radio waves are reflected at a lower altitude than usual, and hence the phase of propagating VLF waves is advanced.

Due to the lack of observation, the exact photon spectrum for each burst was not reported so far. But following \citet{Mereghetti2009} we assume
the spectral shape of an optically thin thermal bremsstrahlung with $kT$=40 keV. 
Then, the number of higher-energy photons increases as the fluence goes up. Higher-energy photons
can penetrate deeper at low altitude and increase the electron number density there. As a result,
the reflection height becomes lower as the gamma-ray fluence increases.

By treating the propagation of VLF radio waves using the mode theory \citep{Wait1964},
we estimated the reduction of the reflection height $\Delta H$ from the phase change $\Delta \phi$.
In the following, we have assumed that the lower ionosphere is isotropic and a sharply bounded medium \citep{Wait1964}, and we have used
a phase velocity expression given by \citet{Wait1959}. Then,
the relation between $\Delta \phi$ and $\Delta H$ can be expressed as \citep{Inan1987}
\begin{equation}
\frac{\Delta \phi}{d \Delta H} \simeq - \frac{2 \pi f}{hc} \left[ \frac{h}{2R_{\rm e}} + C_n^2 \right] \ \ C_n=\frac{ \left( 2n-1 \right) \lambda}{4 h},
\end{equation}
where $d$ is the length of the disturbed region along the great circle path, $f$ is the wave frequency, $h$ is the typical reflection height, 
$c$ is the speed of light, $R_{\rm e}$ is the Earth's radius, $\lambda$ is the wavelength of the VLF radio wave, and $n$ is the order of the waveguide mode.
For a long propagation distance and a normal nighttime reflection height of $h \sim 85$ km, \citet{Wait1964} showed that the second mode ($n$=2) would be dominant.
Therefore, we were able to calculate $\Delta H$ from the observed $\Delta \phi$ for each burst and these values are tabulated in Table 1.
We note that $\Delta H$ calculated by using Eq. (2) is a rough estimate, and Monte Carlo simulations are required to obtain more accurate values.
Nonetheless, this gross estimation would be meaningful to qualitatively consider the effect of gamma-ray illumination.

Next, we consider a mechanism for the observed decrease of VLF wave amplitude during the gamma-ray illumination. We can understand it
on the basis of the altitude dependence of the collision frequency $\nu_{\rm e}$ between electrons and neutral atoms. 
$\nu_{\rm e}$ is higher for lower altitudes, and it is often modeled as $\nu_{\rm e} =1.816 \times 10^{11} \exp(-0.15z)$, where $z$ is the altitude
measured in km \citep[e.g.,][]{Wait1964}. 
Consequently, as the reflection height becomes lower, VLF radio waves are more attenuated, and hence their amplitude decrease.

These hints of correlation suggest that it would be possible to deduce a gamma-ray fluence from $\Delta A$ and $\Delta \phi$.
We note that these relations are applicable only for this particular VLF frequency (21.4 kHz).
There are also another uncertainties which might affect $\Delta A$ and $\Delta \phi$ such as the altitude profile of the ambient electron number density. 
Nonetheless, we claim that the Earth's ionosphere can be used as a new gamma-ray `detector' and VLF data can provide a unique information on 
incident gamma-ray fluences,
even if satellites in space were not able to observe it.
Therefore, we stress here that this VLF method is a new potential technique for monitoring high energy transient phenomena in the universe, once
we know in advance which source is active.

\subsection{Recovery timescale}
We plot in Figure 4 (c) the recovery timescale of each burst against the gamma-ray fluence.
We did not find any significant correlation between the two quantities, and
all the recovery timescales were in the range of 2$-$12 s. 
Since the $t_{\rm rcv}$ are longer than the burst durations (see Table 1), 
the observed VLF amplitude and phase time profiles are different from the gamma-ray light curves.
We also fitted the phase data with a similar function of Equation (1) and calculated the $t_{\rm rcv}$.
Again, we did not find any significant correlation between gamma-ray fluences and recovery timescales.

As shown in Figure 2, recovery time profiles can be well represented by an exponential-like function of a single parameter $t_{\rm rcv}$. 
On the other hand, in the case of magnetar giant flares, two different recovery timescales were reported \citep{Inan2007,Tanaka2008}.
Namely, an initial rapid recovery of a few seconds was followed by a long enduring recovery lasting for $>$1 hour.
The long-duration recovery is interpreted as due to the neutralization of positive and negative ions at an altitude below 60 km \citep{Inan2007},
which means that ionization by gamma-rays mainly occurred at such a low altitude.
In fact, it is known that the photon spectrum and gamma-ray fluence of giant flares are much harder and higher than those of short repeated bursts \citep{Woods2006}.
Lack of such a long-duration recovery in our VLF data suggests that the spectra of short repeated bursts were relatively soft compared to that of giant flares.

As shown above, we interpreted the VLF amplitude and phase changes as due to the lowering of the reflection height. In this case,
faster recovery timescales are expected for larger gamma-ray fluences, because the electron attachment rate
is a negative function of altitude \citep{Rowe1974}. However, as shown in Figure 4 (c), we did not find such a trend. 
This might be due to different spectrum for each burst, contrary to what we have assumed in this paper following \citet{Mereghetti2009}.
The harder the spectrum, the lower the reflection height and the faster the recovery time.
Monte Carlo simulations are required to confirm this possibility, and the results will be reported in a subsequent article.

\section{Summary}
We have detected, for the first time, ionospheric disturbances caused by {\it short repeated} gamma-ray bursts 
from a magnetar. Amplitude and phase changes of VLF propagating waves are correlated with 
gamma-ray fluences. This can be understood in terms of the lowering of the reflection height.
While satellites in space cannot continuously observe the whole sky due to Earth occultation, the Earth's ionosphere can monitor it without
interruption. VLF observations provide us with a new method to monitor high energy transient phenomena of astrophysical importance.

\acknowledgments
Y. T. T. is supported by a JSPS Research Fellowship for Young Scientists.
J. P. R. thanks FAPESP (Proc. 06/02979-0), CNPq (Proc. 304433/2004-7), and MACKPESQUISA funding agencies.

\bibliographystyle{apj}

\clearpage

\begin{figure}
\begin{center}
\epsscale{0.7}
\plotone{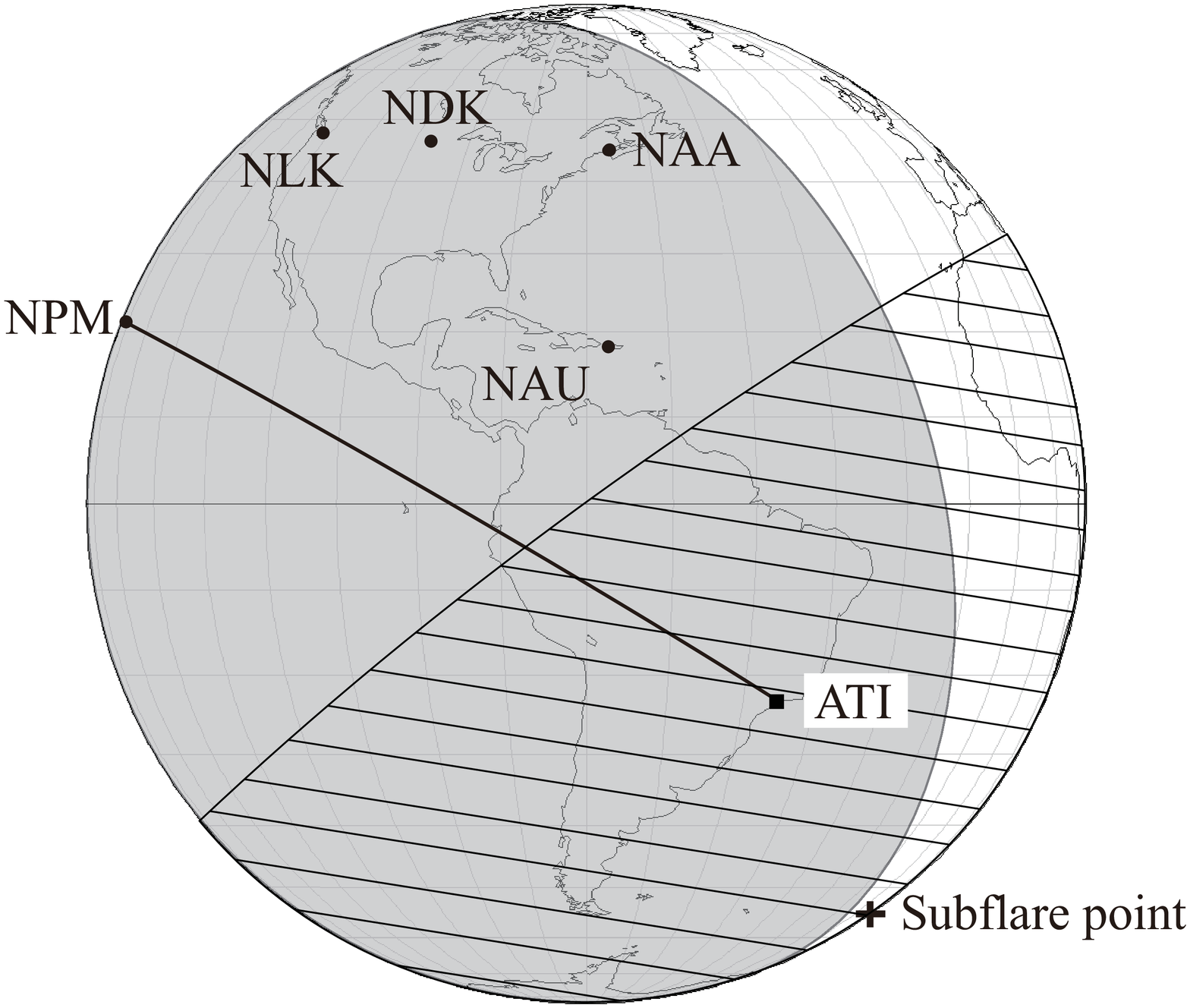}
\caption{VLF propagation path from NPM transmitter (Hawaii) to ATI observing station (S\~{a}o Paulo, Brazil).  
Also shown are the locations of other four VLF transmitters (NLK, NDK, NAA, and NAU).
Shaded hemisphere indicates the night side
part of the Earth at 6:48 UT, when the largest burst occurred (see Table 1). The part of the Earth 
illuminated by gamma-rays at 6:48 UT is also drawn by dashed area.}
\end{center}
\end{figure}

\clearpage

\begin{figure}
\begin{center}
\includegraphics[width=30pc]{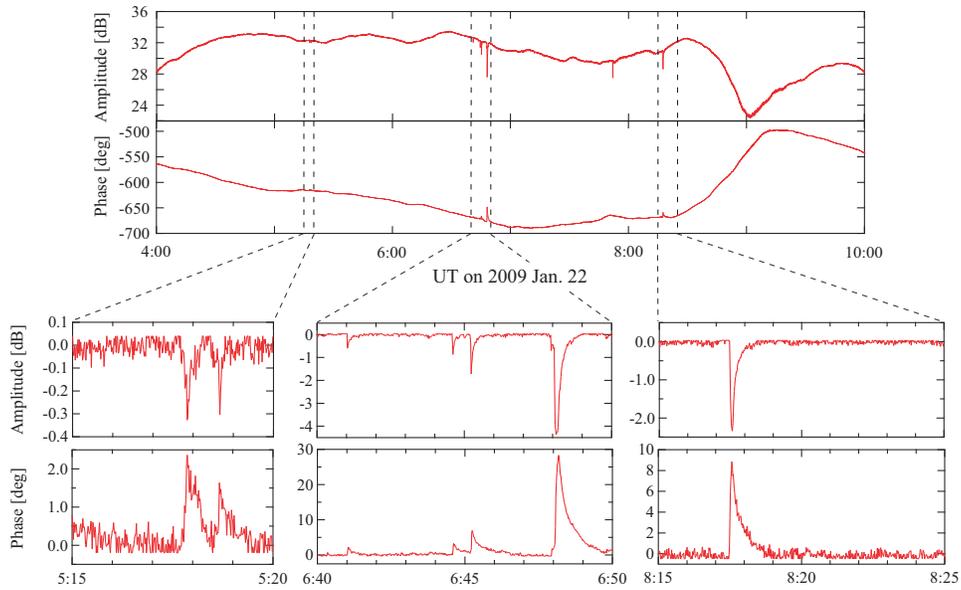}
\caption{Amplitude and phase variations of a VLF signal from NPM transmitter (21.4 kHz), which were observed at ATI (see Figure 1) 
from 4:00 UT to 10:00 UT on 22 January, 2009. Lower figures are background-subtracted blown-ups at time ranges during which short repeated SGR bursts were detected (see also Table 1). 
}
\end{center}
\end{figure}

\clearpage

\begin{figure}
\begin{center}
\includegraphics[width=30pc]{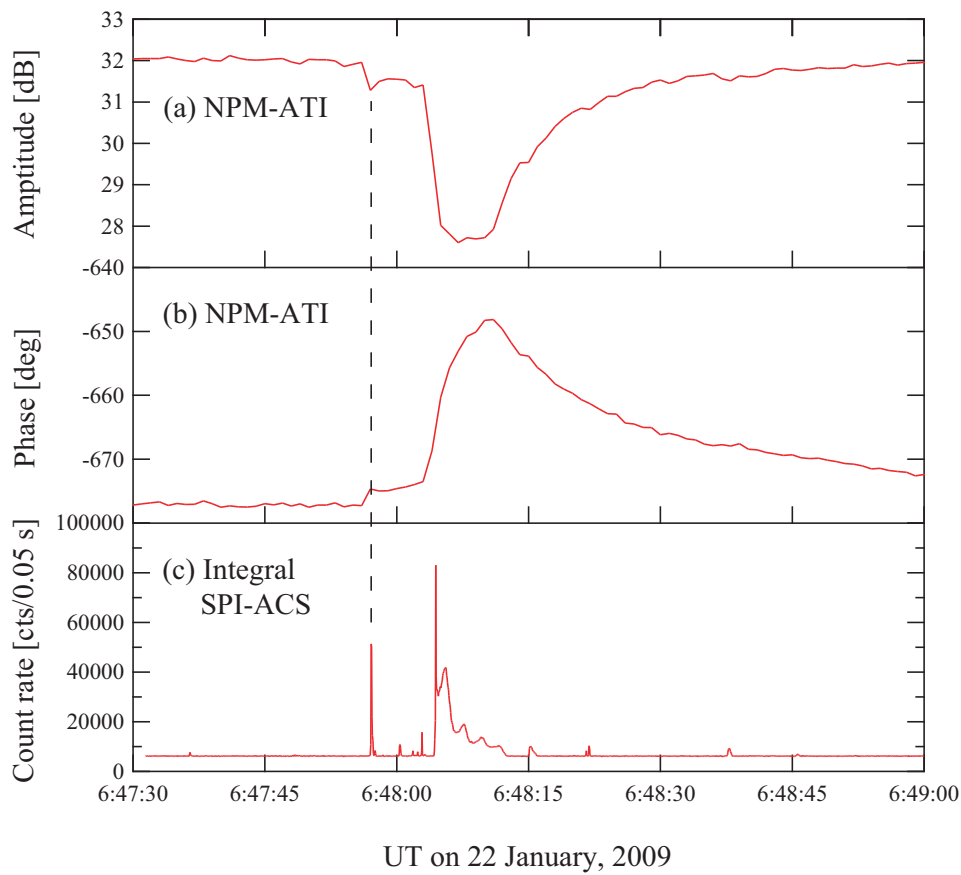}
\caption{(a) Blown-up VLF amplitude data from the NPM-ATI path around 6:48 UT. The vertical dashed line shows the time 6:47:57.1 UT, when a relatively large
gamma-ray flare was observed by {\it INTEGRAL} (see also Table 1). (b) Same as (a), but for VLF phase data. (c) {\it INTEGRAL}/SPI-ACS light curve around 6:48 UT.
Note that the peak of the brightest burst at 6:48:04.3 UT was probably higher than shown here, due to a saturation problem for high count rates \citep{Mereghetti2009}.
}
\end{center}
\end{figure}

\clearpage

\begin{figure}
\begin{center}
\noindent\includegraphics[width=35pc]{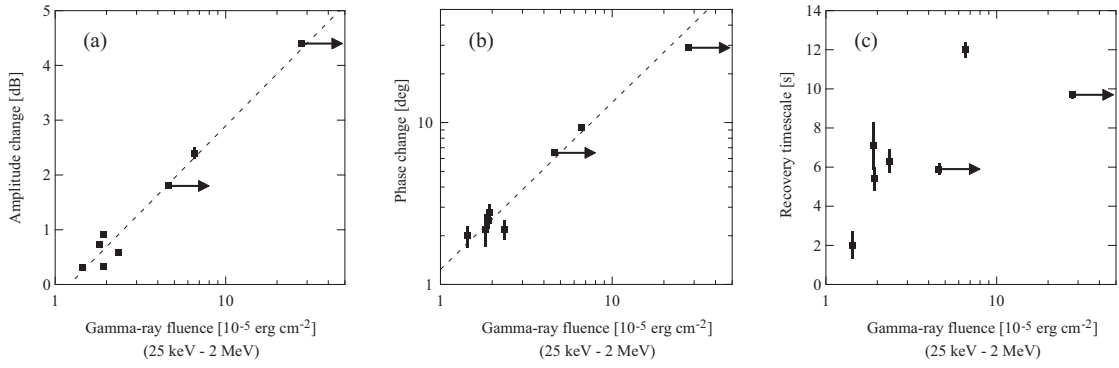}
\caption{(a) Relation between observed amplitude changes and gamma-ray fluences (25 keV $-$ 2 MeV) measured by {\it INTEGRAL}
satellite \citep{Mereghetti2009}. Values are tabulated in Table 1. (b) Same as (a), but for observed phase changes. 
(c) Same as (a), but for recovery timescales estimated by fitting.}
\end{center}
\end{figure}

\clearpage

\begin{landscape}
\begin{table}
\begin{center}
\caption{Short repeated bursts from SGR J1550$-$5418 detected by NPM-ATI VLF data}
\begin{tabular}{ccccccccc}
\hline
Start UT                             & {\it INTEGRAL}       & Duration$^{\dagger}$ & Gamma-ray fluence$^{\dagger}$$^{\ddagger}$ & Amplitude    & Phase             & Recovery timescale & Lowering of\\ 
on 22 Jan.$^{\dagger}$ & ID$^{\dagger}$ & [s]                                    & 10$^{-5}$ [erg cm$^{-2}$]              & change [dB] & change [deg] & of amplitude [s] & reflection height [km]\\ \hline
5 17 51.7                          & 85   & 0.45                                & 1.91$\pm$0.01                                 & $-0.33\pm0.03$ & 2.5$\pm$0.3 & 7.1$\pm$1.2 & 8.0$\pm$1.0\\
5 18 39.5                          & 93   & 1.00                                & 1.44$\pm$0.01                                 & $-0.32\pm0.03$ & 2.0$\pm$0.3 & 2.0$\pm$0.7 & 6.4$\pm$1.0\\
6 41 02.1                          & 108 & 1.00                                & 2.35$\pm$0.01                                 & $-0.59\pm0.04$ & 2.2$\pm$0.3 & 6.3$\pm$0.6 & 4.5$\pm$0.6\\
6 44 36.4                          & 117 & 1.75                                & 1.93$\pm$0.01                                  & $-0.91\pm0.05$& 2.8$\pm$0.3 & 5.4$\pm$0.6 & 5.8$\pm$0.6\\
6 45 13.9                          & 121 & 1.45                                & $>$4.59$\pm$0.01                           & $-1.8\pm0.05$  & 6.5$\pm$0.3  & 5.9$\pm$0.3 & 13$\pm$0.6\\
6 47 57.1                          & 141 & 0.35                                & 1.82$\pm$0.01                                  & $-0.74\pm0.04$ & 2.2$\pm$0.5 & $-^{\ast}$ & 4.5$\pm$1.0\\
6 48 04.3                          & 149 & 8.15                                & $>$27.76$\pm$0.03                         & $-4.4\pm0.04$   & 29$\pm$0.5 & 9.7$\pm$0.2 & 60$\pm$1\\
8 17 29.4                          & 176 & 6.20                                & 6.59$\pm$0.02                                  & $-2.4\pm0.1$ & 9.3$\pm$0.3 & 12.0$\pm$0.4 & 14$\pm$0.5\\ 
\hline
\end{tabular}
\end{center}
$^{\dagger}$ Taken from \citet{Mereghetti2009}.
$^{\ddagger}$ The energy range is 25 keV to 2 MeV.
$^{\ast}$ Unable to determine by fitting.
\end{table}
\end{landscape}

\end{document}